\documentclass[aps,prstper,10pt]{revtex4-2}

\usepackage{multirow}
\usepackage{graphicx}% Include figure files
\usepackage{dcolumn}% Align table columns on decimal point
\usepackage{bm}% bold math
\usepackage{tabularx}
\usepackage{booktabs}
\usepackage{hyperref}

\hypersetup{
    colorlinks=true,
    linkcolor=blue,    % color for internal links
    urlcolor=blue,      % color for external links
    citecolor=blue     % color for citations
}

\begin{document}

%\preprint{APS/123-QED}
\title{Assessing Students' Understanding of Uncertainty in Undergraduate Physics Laboratory Courses at a Major Canadian University}

\author{Matheus A. S. Pess\^oa}
\email{matheus.pessoa@mail.mcgill.ca}
\affiliation{McGill University Physics Department}
\affiliation{Office of Science Education}
\author{Rebecca Brosseau}
\affiliation{McGill University Physics Department}
\affiliation{Office of Science Education}
\author{Benjamin J. Dringoli}
\affiliation{McGill University Physics Department}
\author{Jack C. Sankey}
\affiliation{McGill University Physics Department}
\author{Thomas Brunner}
\affiliation{McGill University Physics Department}
\author{April Colosimo}
\affiliation{McGill Libraries}
\author{Armin Yazdani}
\affiliation{Office of Science Education}
\author{Janette Barrington}
\affiliation{Office of Science Education}
\author{Kenneth Ragan}
\affiliation{McGill University Physics Department}
\author{Marcy Slapcoff}
\affiliation{Office of Science Education}
\date{\today}% It is always \today, today,
             %  but any date may be explicitly specified

\begin{abstract}
An article usually includes an abstract, a concise summary of the work
covered at length in the main body of the article. 
\begin{description}
\item[Usage]
Secondary publications and information retrieval purposes.
\item[Structure]
You may use the \texttt{description} environment to structure your abstract;
use the optional argument of the \verb+\item+ command to give the category of each item. 
\end{description}
\end{abstract}

%\keywords{Suggested keywords}%Use showkeys class option if keyword
                              %display desired
\begin{abstract}
Over the last five years, McGill University’s Office of Science Education (OSE) has partnered with faculty members from the Department of Physics to form an education research group with the aim of charting the progression of students’ conceptual understanding of uncertainties across their undergraduate degree. The research conducted by this group seeks to provide further insight into the experimental skillset that students gain through undergraduate laboratory courses and noticeable gaps in student understanding that the department could address. In this paper, we evaluate the conceptual understanding of uncertainty using the Concise Data Processing Assessment (CDPA) test. First, we characterize the physics laboratory curriculum at  McGill University by evaluating the evolution of CDPA scores across consecutive laboratory courses, and further propose the utilization of this tool for identifying gaps in student understanding. Following the analysis of student responses (N = 2023) we specifically investigate data collected in second-year courses to better understand what student errors reveal about common misconceptions in experimental physics. This more in-depth analysis focuses on data collected from students at the
beginning and end of two consecutive experimental laboratory courses that build on each other. By the end of the second course, students have engaged with all the material covered in the CDPA test. Overall results show an upward shift in student understanding of uncertainties over time.  Interestingly, we did not observe any change in CDPA scores comparing throughout and post pandemic scores. Despite the upward shift, many students continue to struggle with uncertainties, basic data analysis and curve fitting. 
\end{abstract}
\maketitle
\section{Introduction}
Understanding measurements of physical quantities and uncertainties is an important skillset for evaluating and interpreting data in scientific research. In recent years, the consideration and discussion of uncertainties has increasingly become more present and relevant in physics undergraduate curricula \cite{wan2023investigating, stump2023comparing, may2023historical}, in line with the American Association of Physics Teachers (AAPT) recommendations for undergraduate physics labs \cite{kozminski2014aapt}. Skills such as modeling \cite{brewe2008modeling}, experimental design \cite{karelina2007acting}, data analysis, interpretation \cite{pillay2008effectiveness}, and communication \cite{rodriguez2012communicating} are essential for the training of physicists. Developing an understanding of where the competencies are not being addressed within the physics curriculum can help professors develop pedagogical practices to address these gaps.

The study reported in this paper was conducted at McGill University, a research-intensive university in Quebec, Canada. Following an in-depth curriculum review process, the Chair of Physics Department invited a small group of professors teaching at different levels in the experimental physics program to partner with the University’s Office of Science Education (OSE) to form the physics education research (PER) group. The focus of the research was chosen to be improving introductory, intermediate, and advanced undergraduate physics labs to reinforce the understanding of uncertainties was selected the focus of our research. This focus emerged through in-depth discussions about assessment conducted at peer institutions \cite{kung2006university, day2011development, lewandowski2018initial, holmes2020investigating, geschwind2024using}. These studies highlight the importance of assessing the understanding of uncertainties as a means to revise and improve students' experimental competencies in first-year laboratory courses \cite{pollard2017impact, lewandowski2017student}. 

In the PER literature, a range of tests are used to assess experimental competencies such as the Concise Data Processing Assessment (CDPA) \cite{day2011development}, the Colorado Learning Attitudes about Science Survey for Experimental Physics (E-CLASS) \cite{wilcox2016students}, the Physics Measurement Questionnaire (PMQ) \cite{campbell2005teaching}, the Laboratory Data Analysis Instrument (LDAI)\cite{eshach2016developing}, the Physics Lab Inventory of Critical Thinking (PLIC) \cite{walsh2019quantifying}, and more recently the Survey of Physics Reasoning on Uncertainty Concepts in Experiments (SPRUCE) \cite{vignal2023survey}. Each of these evaluates different aspects of physics students' learning. PLIC, for example, focuses on students' evidence-based decision making by measuring abilities ranging from evaluation of data, methods, and conclusions to the proposal of next steps in hypothetical experiments. In comparison, SPRUCE and CDPA are examples of specific tools to assess the understanding of the concept of uncertainty in undergraduate labs. SPRUCE has a broader approach, identifying sources of uncertainty while an experiment is being conducted and leading up to the handling of uncertainty, whereas the CDPA is used specifically to assess error propagation skills, and the interpretation of equations, graphs, and uncertainties therein.

The CDPA was chosen by the McGill PER group as the primary investigative tool to conduct a targeted, longitudinal study across the physics department's undergraduate laboratory courses. The objective was to chart the progression of students' understanding of uncertainty over the course of their degree. In this paper, we present the results of a 4-year study in which we implemented the CDPA test across six courses, with a large dataset of N=2023 responses to the CDPA. Through this research, we focus on discussing relevant findings in fundamental PER topics that serve to guide the department’s current teaching goals and future pedagogical directions.

The objectives of this work are 1) to establish a validation of baseline CDPA data at McGill University compared to the data published in a previous study \cite{day2011development}, 2) to investigate the progression of CDPA scores throughout different lab courses in consecutive years, 3) to compare the evolution of student comprehension on the topic of uncertainty prior to, during, and directly following the global COVID-19 pandemic period (2019-2022) during which students experienced varying degrees of virtual versus in-person instruction, and finally 4) to propose a deeper interpretation of CDPA results in terms of mapping student abilities and identifying common student misconceptions. Where possible, we also provide additional contextual explanations of the features of our findings that may influence the data interpretation.

\section{Background}

The CDPA was first implemented in 2011 at the University of British Columbia (UBC) and the University of Edinburgh \cite{day2011development}. It was initially developed for first-year physics major students in laboratory courses or in science-intensive programs that also required an experimental physics laboratory component. Graduate students and professors also participated in the study in order to ensure the validity of the test in measuring novice vs. expert performance and to further investigate its reliability when used by students at different levels of instruction. The CDPA test consists of 10 multiple choice questions, which vary in terms of difficulty, each with four possible answers. The questions cover a variety of learning outcomes such as identifying the adequacy of a fit to a given dataset, handling instrumental uncertainty, and inferring measurement units from the data. The test’s characteristics, such as its difficulty and its capacity to discriminate amongst the level of expertise of test-takers, are proposed and quantified in the original paper. Furthermore, the test is administered both as a pre-course and post-course assessment in order to reliably measure shifts in reasoning that can be linked to the learning outcomes of each course. 

The results in \cite{day2011development} showed 1) that there are positive shifts in pre- to post- comparisons for the first-year courses, 2) that the test is sensitive to the level of expertise of the test-taker, and thus is able to distinguish between undergraduate students, graduate students, and professors, 3) that the test has wide applicability, a conclusion made based on the comparable results from  UBC and the University of Edinburgh. All these findings are supported by statistical tests which confirm the reliability of the CDPA. For example, first-year UBC students scored an average of $3.91 \pm 0.12$ (standard error of the mean, SEM) in the post-test, compared to an average of $5.31 \pm 0.42$ (SEM) for the graduate students, and $8.00 \pm 0.35$ (SEM) for professors. A similar, more marginal progression of knowledge was also shown across the undergraduate program by comparing the post-test of first-year students with that of fourth-year students who averaged $4.73 \pm 0.26$ (SEM) in the assessment.

Since its publication, the CDPA has stimulated further research within the PER community with different applications of the CDPA results. For example, \citet {day2016gender} investigated the influence of gender in the assessment results, finding no evidence of gender differences in conceptual understanding of uncertainties. In addition,  \citet{holmes2014making} used the CDPA to investigate the role of guidance in productive failure  during the learning process. Furthermore, \citet{kontro2019development}  explored the use of the CDPA for intermediate-level undergraduate physics courses and proposed the possibility of a relationship between CDPA final scores and E-CLASS results, though no significant correlation is observed.

\section{Methodology}
\subsection{Context}
The research conducted for this study took place between 2019 to 2023. This period included the global pandemic in which instructors were forced to adopt new pedagogical strategies for the online environment, while  maintaining course content and learning outcomes. When the pandemic hit in 2020, courses moved completely online. In 2021, a hybrid format was adopted, in which lectures were held online and lab sessions were mostly delivered in person. During 2022, lectures and labs transitioned back to a fully in-person format. In keeping with the CDPA requirements, no adaptations were made to the original CDPA test and the entire ten-question test was provided to all student respondents. Furthermore, as requested by the authors, we did not reproduce or include direct citations of any of the ten questions in order to preserve the use of this instrument for future research. To incentivize participation, a bonus of $1\%$ was offered to all students in each class if at least $80\%$ of students completed the CDPA test.

Finally, to avoid statistically significant shifts due to students retaking the CDPA test from one year to the next, we asked student respondents to identify whether they had taken the CDPA test in a previous semester. The test was conducted online through MS Forms and student respondents were instructed to use up to 30 minutes to complete it. The pre-tests were taken over the first three to four weeks of term, whereas post-tests were taken during the last three to four weeks of term. Tests without pre and post distinction were taken at the end of the term. We collected data in 8 courses: PHYS 101- Introduction to Mechanics, PHYS 131- Mechanics and Waves, PHYS 257- Experimental Methods I, PHYS 258- Experimental Methods II, PHYS 359- Honors Laboratory in Modern Physics I, PHYS 439- Majors Laboratory in Modern Physics, PHYS 469- Honors Laboratory in Modern Physics 2, and PHYS 434- Optics. All results presented here are summarized in Appendix \ref{appendixtables}.

\subsection{McGill Physics curriculum}
The understanding of uncertainty was identified by professors as a core concept in the undergraduate Physics laboratory curriculum. This is reflected in the increasing weight given to assessments that measure this concept as students move from introductory to advanced laboratory courses.

At McGill University, there are different streams for students taking lab courses: first-year students in the general science program (or those pursuing a minor in physics) are required to take an introductory undergraduate laboratory course, PHYS 101- Introduction to Mechanics; physics majors and honors students take a higher-level course PHYS 131- Mechanics and Waves. In both these courses, students explore basic mechanics concepts through simple confirmation experiments where the concept of uncertainty is addressed but does not constitute a major learning outcome. Students who proceed to a physics-specific degree, i.e., majors and honors students, enroll during their second year in PHYS 257- Experimental Methods I and PHYS 258- Experimental Methods II, where they are formally introduced to uncertainties in experiments. In these two more advanced courses, students are required to report uncertainties and discuss the validity of their experimental results and findings based on their interpretation of these uncertainties.

By the conclusion of PHYS 257 and PHYS 258, students are expected to possess the skills needed to interpret, address, and visually represent measurements. Students who choose to focus on experimental physics take the 300- and 400-level courses, which are not part of the mandatory curriculum, and can therefore represent expertise-like level of understanding. The notation difference between 300 and 400 is simply a means to indicate which students are part of the honors program as opposed to the majors program, otherwise the curricular content covered in both PHYS 359 and PHYS 439 is identical. We combined the results for advanced courses (PHYS 434, PHYS 439 and PHYS 469) to form the overall 400 level, because the sample size of each class was too small and the students' level of expertise was similar. 

\subsection{Educational background}
In the context of our university, situated in Québec, a notable portion of incoming freshman students exhibit diverse educational backgrounds. This diversity is attributed to the fact that some students have completed Collège d'Enseignement Général et Professionnel (CEGEP) before enrolling in university. Students who complete CEGEP and pursue a physics degree at univeristy skip the first year of the McGill curriculum, taking PHYS 257 and PHYS 258 as their initial university-level laboratory courses. During data collection and analysis, we could not differentiate whether students had completed CEGEP or not.

\subsection{Physics Education Research (PER) Group at McGill}
\subsubsection{Origin of the PER group}
The PER group leading this study represented a partnership between the Department of Physics and the Office of Science Education (OSE). Following an in-depth curriculum review process, the chair of the department nominated five tenure track professors teaching at different levels in the experimental physics program to work with OSE staff (including the director, educational developers and a science education expert),  the physics liaison librarian, as well as PhD students selected by the department. The goal was to measure the effectiveness of lab courses from a program perspective. Although membership evolved over the five-year period, the group met on a regular basis and operated similarly to a community of practice \cite{cuddy2002cultivating} with distributed leadership and regular cycles of data collection and reflection.
 
\subsubsection{PER professors feedback sessions}
Following the most recent round of data collection, feedback sessions were held with the PER professors. The purpose of these sessions was to better understand the level of each department’s experimental courses and to map the specific skills taught in each course as they related to the content of the CDPA questions. These sessions were used to situate each course in the McGill Physics undergraduate program based on their content, and to better understand how students at various points in their degrees would be expected to respond to the CDPA. This analysis will be discussed further below. 

\subsection{Statistical analysis}
Shapiro-Wilk \cite{shapiro1965analysis} tests and Kolmogorov-Smirnov \cite{massey1951kolmogorov} test revealed non-normality in the distribution of CDPA scores, parametric tests were not suitable for between-group comparisons. Therefore, non-parametric tests were employed. Between-group differences in CDPA scores were assessed for statistical significance using the Kruskal-Wallis \cite{kruskal1952use} tests. Pairwise comparisons were conducted to compare all groups and a $p$ value $<0.05$ was considered as statistically significant. For multiple comparisons, Dunn’s post hoc comparison test \cite{dunn1961multiple} was applied to adjust the $\alpha$-level as necessary. Bonferroni \cite{bonferroni1936teoria} correction was not applied. All statistical analyses were performed with GraphPad Prism ®(GraphPad Software Inc., San Diego, Calif., USA). All the graphs shown in this article display the mean as well as the standard error of the mean.

\section{Results and Discussions}
%\subsection{Initial findings in departmental-wide analysis}
%\subsubsection{Progression as a function of levels of expertise}

\subsection{Comparison to UBC}
\begin{figure*}
  \centering  \includegraphics[width=1\textwidth]{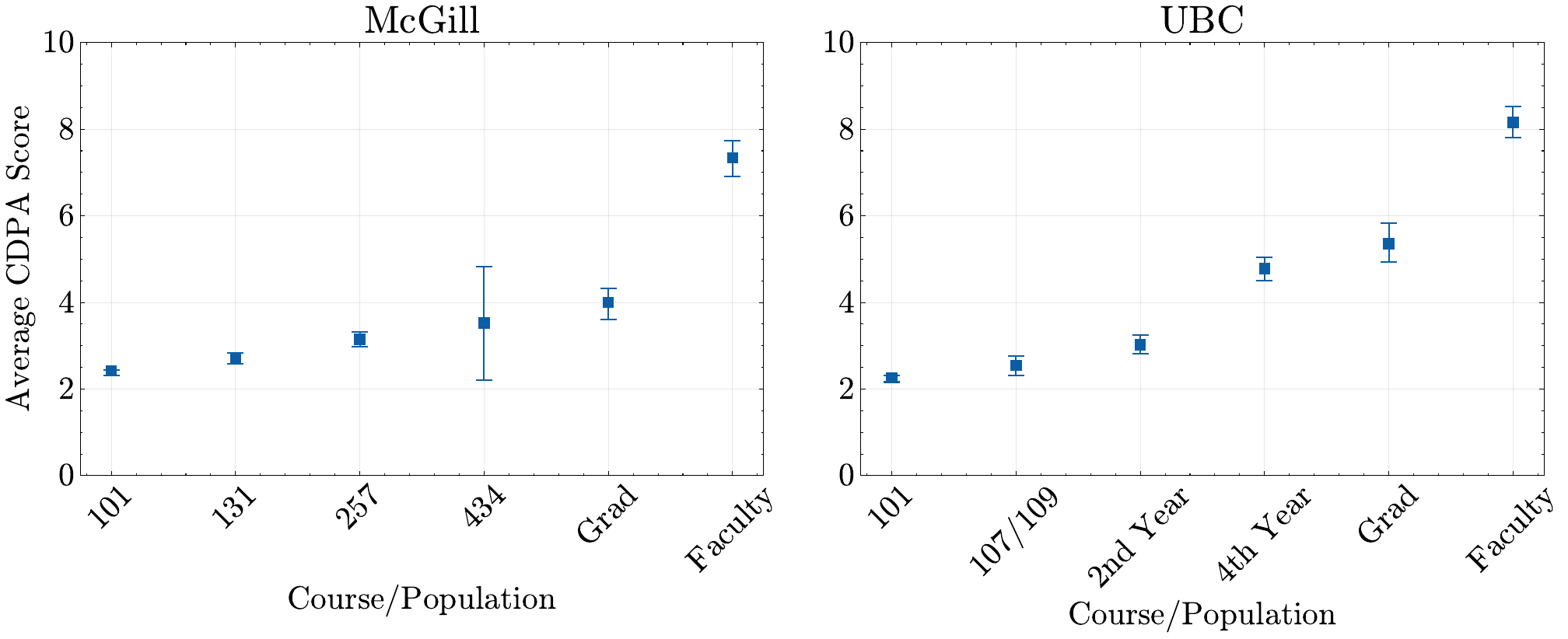}
  \caption{Mean CDPA scores (left) for McGill University and (right) the University of British Columbia (UBC), as a function of level of program/degree stage in the year of 2019. The average score increases as a function of level of instruction.}
  \label{fig:ubccomparison}
\end{figure*}

To validate the CDPA tool, in 2019 we decided to compare McGill's CDPA results to those at University of British Columbia (UBC), where the tool was first developed. In Figure \ref{fig:ubccomparison}, we present the results at the two universities. The results demonstrate a similar difference between novice and expert conceptual understanding of uncertainties at both universities. We notice a maximum score for students in the fourth and final year of their undergraduate education, without significant change towards the graduate student level. As expected, professors from both universities score higher than all previous cohorts, but not a perfect score. These CDPA results allow us to draw an important conclusion: even at the higher end of the novice to expert scale, the average score is not a complete $10$, a feature consistent with Day et. al. \cite{day2011development}. 
 
 %This reinforces how a score for undergraduate students, which is between 2.0 and 5.0 is not necessarily indicative of instructional content that is missing from the McGill Physics curriculum.
 
\subsection{Longitudinal analysis of the McGill data}

\begin{figure*}
  \centering
  \includegraphics[width=\textwidth]{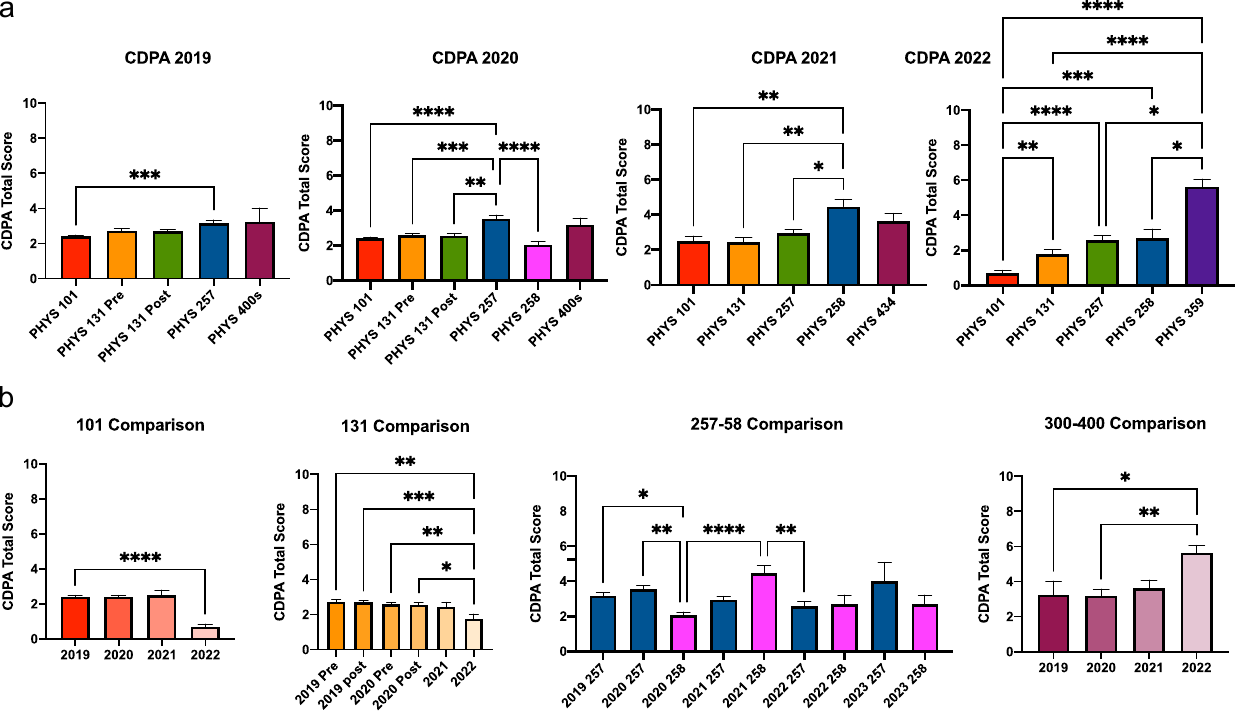}
  \caption{(a) Bar graphs representing the total CDPA score out of 10 for first year (PHYS 101, PHYS 131), second year (PHYS 257 and PHYS 258), and third year and upper-level (PHYS 359, PHYS 400s, PHYS 434) undergraduate laboratory courses at McGill University, across four years (2019-2922). The stars (*) represent the degree of statistical significance between courses, and the errorbars refer to the standard error on the mean (SEM). (b) Total CDPA score out of 10 for the undergraduate laboratory courses, presented individually as a function of year.}
  \label{fig:yearyear}
\end{figure*}

The total CDPA mean score for the courses investigated between 2019-2023  is presented on a scale from 0 to 10 in Figure \ref{fig:yearyear}a. From left to right, the order of level of expertise increases by course numbers. We notice statistically significant jumps in the CDPA scores when comparing  the scores obtained during the very first mechanics course that contains a lab component, PHYS 101, to a more advanced laboratory, PHYS 257. This can be further explained by the fact that in PHYS 257 students are more equipped to deal with uncertainties and apply methods to analyze them experimentally.

Moreover, in 2019 and 2020, we also analyzed pre- and post-test results for the initial first-year mechanics course, PHYS 131, to evaluate changes within the course because, throughout this course, students are asked to calculate uncertainties but are not specifically evaluated on that skill. In these two years, we notice that the CDPA scores obtained for the pre- and the post-test are similar, which is to be expected since students are only exposed to uncertainties through problem-solving. When comparing PHYS 131 with higher level courses, however, we do notice a statistically significant increase in the level of responses when compared to the experimental course PHYS 257 in 2019 and in 2020. In 2021, when only PHYS 131 post-test data was collected, we also notice a statistically significant jump between PHYS 131 and PHYS 258, which marks the increase in learning following undergraduate students' second full year of experimental courses.

When comparing PHYS 257 to PHYS 258, we notice an increase in student understanding that is consistent across three out of four years under investigation. Most notably, however, the average CDPA score collected from students at the end of PHYS 359 is significantly superior to the results from all previous courses, whereas, for the 400-level courses, because the number of student respondents was smaller and their answers presenting a higher degree of deviation, we cannot make of the same claims.

%\subsubsection{Consistencies within courses}
In Figure \ref{fig:yearyear}b, we compare how the CDPA scores evolve for the same course as a function of time. Interestingly, for PHYS 101 and PHYS 131, the overall CDPA score has a combined average that does not change as a function of year, except in the rare case of PHYS 101 in 2022 where the score was abnormally low. It is important to note that no major changes were implemented in the course that could have explained this difference in score. In PHYS 257 and PHYS 258, the results show fluctuations that can be consistent with the learning curve faced at that level for each of the different cohorts. The structure, content and evaluation methods used in these two courses did not change in any of the years under study. However, different professors taught the course each year using the same syllabus. In the 2020 Winter session of PHYS 258, due to the pandemic, students only completed 60 \% of the course -- the laboratory component. This may partially explain the decrease in CDPA scores that year. 

For higher level courses over the first three years, the overall score is also consistent with a jump observed in 2022. In that year, the number of students taking part in the class was less than in previous years but students still obtained significantly higher scores than previous cohorts. These findings are likely due to classroom dynamics that are notably different from the 200 level courses, which we will further argue plays an important role in students' understanding of uncertainty. 

%For the different years and formats (pre and post), the overall CDPA scores are consistent within the levels of instruction of each course. 

%These scores correspond to an average that is even lower than expected for random guessing, and we cannot pinpoint the reason for that, since no major changes have been implemented in the course. For the different years and formats (pre and post), the overall CDPA scores are consistent within the levels of instruction of each course. 

%In PHYS 257 and PHYS 258, we observe fluctuations but with an overall increase in the total score towards PHYS 258 comparing the pairs of years corresponding to a full academic year. 

%The higher-level courses present less fluctuations in the overall score

The data presented in Figure \ref{fig:yearyear} shows that for the two lower-level courses, PHYS 101 and PHYS 131, the overall CDPA score can be understood as being comparable to random guessing. This can be explained by the fact that, in those courses, students are introduced to the concept of uncertainties for the first time and are only capable of applying simple mechanics problems and experiments without a deeper conceptual understanding. The CDPA test evaluates student understanding in topics that involve uncertainty propagation, graph analysis with fits and fitting uncertainties, logarithmic fits, and evaluation of $\chi^2$. In the original CDPA paper \cite{day2011development}, the authors discuss that the vast majority of students from introductory-level laboratory courses are not yet ready to answer all the survey questions in the first year of their undergraduate degree. These results are in agreement with our observations and analysis of the understanding of uncertainty for students enrolled in PHYS 101 and PHYS 131.

In the following discussion, we hone in on the results obtained from the PHYS 257 and PHYS 258 courses because it is during this full academic year that students learn how to perform curve fitting, start considering uncertainties to assess the outcome of an experiment, and produce and interpret graphs based on measuring uncertainties. Comparisons at the higher level could also garner interesting results, however the population size of student respondents was not sufficiently large for each year of the study. Additionally, having pairwise comparisons between two courses within the same academic year allows for a better comparison in terms of the understanding of uncertainties. In the case of the McGill Physics curriculum, the two courses are complementary to one another and students' abilities to understand concepts related to uncertainty is not developed until the end of the full academic year, as shown in Table \ref{table:abilities}. At the end of PHYS 258, students are asked to conduct a small research project in which they need to answer a research question of their choice. Students experimentally validate their measurements and study how the uncertainties should be considered in the analysis, as well as how to best mitigate errors.

In Table \ref{table:abilities}, we present the central concepts regarding uncertainties that are covered in each of the courses we investigate. While in PHYS 101 students know how to propagate uncertainties numerically and make graphs, it is not until PHYS 257 and PHYS 258 that they actually get to do more advanced experiments in which error analysis becomes crucial. In these two courses, the students' experimental skills are put to the test and they are evaluated based on applications involving uncertainty propagation at different levels. In PHYS 257, students are exposed to and are required to assess systematic errors and perform error propagation. Moreover, in PHYS 258, they need to consider statistical uncertainties to get the best-fit curves, understand and justify their reasoning, and differentiate between different models. We present in Table \ref{table:example_questions} some example questions from the laboratory manuals at three different course levels.

\begin{table*}[ht]
\centering
\renewcommand{\arraystretch}{1.3} % Adjust row height
\resizebox{\textwidth}{!}{%
\begin{tabular}{l>{\centering\arraybackslash}m{2.5cm}>{\centering\arraybackslash}m{2.5cm}>{\centering\arraybackslash}m{3.0cm}>{\centering\arraybackslash}m{2.5cm}>{\centering\arraybackslash}m{2.5cm}>{\centering\arraybackslash}m{2.5cm}>{\centering\arraybackslash}m{3.0cm}>{\centering\arraybackslash}m{3.0cm}}
\toprule
\textbf{Course} &
  \textbf{Uncertainty propagation} &
  \textbf{Experiments} &
  \textbf{Analysis of uncertainties} &
  \textbf{Uncertainty minimization} &
  \textbf{Graph construction} &
  \textbf{Graphs with uncertainties} &
  \textbf{Grade for uncertainties} &
  \textbf{Experimental design} \\ 
\midrule
PHYS 101 & x & x  &   &   & x &   & x  &   \\ 
PHYS 131 & x & x &   &   & x &   & x  &   \\ 
PHYS 257 & x & x & x & x & x & x &  x &   \\ 
PHYS 258 & x & x & x & x & x & x & x & x \\ 
PHYS 359 & x & x & x & x & x & x & x & x \\ 
PHYS 400s & x & x & x & x & x & x & x & x \\ 
\bottomrule
\end{tabular}
}
\caption{Evaluation criteria related to uncertainties for each course in the study. Notably, since the first course, students are exposed to uncertainties and experiments. However, it is only at higher level laboratories with experiments that require a higher degree of precision that students need to evaluate the experimental uncertainties. }
\label{table:abilities}
\end{table*}

\begin{table*}[ht]
\centering
\renewcommand{\arraystretch}{1.3} % Adjust row height
\begin{tabular}{p{3cm} p{12cm}} % Adjust column width as needed
\toprule
\textbf{Course} & \textbf{Example Questions} \\
\midrule
PHYS 101, 

PHYS 131 & 
\begin{itemize}
  \item The value of the mass used in the experiment (measured with a scale) and its uncertainty.
  \item Discuss the uncertainty estimates. How did you measure uncertainties?
\end{itemize} \\
\midrule
PHYS 257 
& 
\begin{itemize}
  \item Compare the standard deviation obtained from the manual measurements with the automated measurements made. Which measurement is more reproducible/precise? Comment on the difference. Assuming that systematic errors are absent, what effect does this have on the uncertainty in your measurement? Be quantitative in your answer, not vague and qualitative. 
\end{itemize} \\
\midrule
PHYS 258 & 
\begin{itemize}
  \item For data analysis, we will convert our data tables into histograms. We will then fit Gaussian and Poissonian distributions to the histogram and compare $\chi^2$ and residual plots with each other.
\end{itemize} \\
\bottomrule
\end{tabular}
\caption{Example instructions taken from a laboratory manual of each lab. The treatment of uncertainties is mostly numerical, without a critical assessment. This changes in higher level labs, where students are asked to identify, interpret, and mitigate possible uncertainties faced experimentally.}
\label{table:example_questions}
\end{table*}

%- These results can tell us more than just the percentage of the correct responses, and we are interested in that.

%- Why is CDPA more applicable to 257 and 258 first-year lab courses; It`s something that is motivated by the findings we present here but also a finding of Bonnn 2011 (original CDPA paper), not all students in 101 and 131 are ready to respond all questions

%- one clean full year of data is from students when they are exposed to the material; the cleanest way to assess the misconceptions

%- Deciding the most important questions (this is where we talk about the professor interaction); based on original CDPA paper, 4 questions were identified as the most important for the conceptual understanding of uncertainty (1,2,5,6,9,10). For the purposes of our study we will be focusing on 1 6 and 10 for the 257 and 258 cohort

\subsection{Impact of the pandemic }

The global COVID-19 pandemic of 2020 introduced a significant external factor that we were able to study. Before, during, and after the pandemic the content of the courses included in this study did not change. However, the delivery formats changed due to regulatory health guidelines. Once smaller courses returned to an in-person teaching environment, many of the large-scale introductory courses, including PHYS 101, PHYS 131, PHYS 257, and PHYS 258 continued to be taught in either a hybrid or a fully remote format. 

\begin{figure*}
  \centering
  \includegraphics[width=0.5\textwidth]{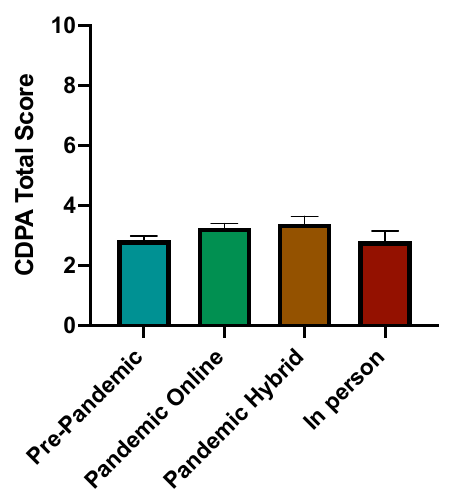}
  \caption{CDPA average scores for PHYS 257 and PHYS 258, combined, in four periods: 1) pre-pandemic, 2) pandemic (with online classes), 3) pandemic hybrid (theoretical classes online, laboratory classes in -person), and 4) in person.}
  \label{fig:pandemic}
\end{figure*}

Surprisingly, we notice no significant difference in CDPA scores between these pre- and post-pandemic periods. As was pointed out by one of the course professors, this lack of any statistically significant fluctuation in CDPA scores is, in and of itself, noteworthy. The consistency of these scores, as shown in Figures \ref{fig:yearyear} and Figure \ref{fig:pandemic}, suggests that the pedagogical benefit of laboratory courses may have more to do with students gaining first-hand practical knowledge through manipulating laboratory instruments than with their conceptual understanding of measurement principles. 

Furthermore, if exposure to experimental instruments is only part of the intended pedagogical value of laboratory courses, then how can these in-person labs be improved such that they enhance students' understanding of concepts related to uncertainty to a degree that is distinguishable using a measurement instrument such as the CDPA? The consistency of these pandemic-era CDPA scores also raises the question whether the CDPA instrument can distinguish between the impact of confirmation versus inquiry-guided instructional approaches. From the findings of this research, we deduce that in all undergraduate courses included in this study could benefit from more dedicated teaching of graphing and analysis of uncertainties, as we will discuss in the following sections.

\subsection{Assessing student misconceptions}

Wilson et al. \cite{wilson2010measurement} and Harrison et al. \cite{harrison2017threshold} describe the concept of uncertainty of physical measurements as a threshold concept in physics, emphasizing its critical role in deepening students' understanding of various topics, including experimental data analysis and the fundamental principles of quantum physics \cite{majidy2024addressing}. Majidi's review article further highlights that misconceptions about uncertainties can arise from three primary sources: the knowledge itself, teaching methods, or the students' perspectives. Overall, misconceptions can be more detrimental than gaps in knowledge. This is because students tend to resist new information that conflicts with their existing beliefs, resulting in cognitive dissonance and frustration. We can use the CDPA to identify misconceptions, acknowledging that we are limited to assessing the alternatives present in the test. 

Furthermore, Wilson et.al \cite{wilson2010measurement} discuss that understanding uncertainties can be divided into three different layers of understanding: pattern recognition, formal procedure, and meaning. In this study, three skills were evaluated through interviews with students, and it was discussed that integrating the three elements can give a more complete view on student understanding. In terms of the CDPA, pattern recognition is applicable to questions that evaluate uncertainties in graphs, or numerical uncertainties. The second skill, formal procedure, is tested when students have to actually calculate the uncertainties for some of the numerical questions presented in the assessment. The skill related to the meaning of uncertainties cannot be directly assessed through the CDPA, but can potentially be inferred through the experimental outcomes of the laboratory activities. 

By analyzing the results from novice to expert levels in two Canadian universities (Figure \ref{fig:ubccomparison}), we see that the CDPA is not an easy test. The distractors represented by each of the incorrect answers, which are linked to distinct numerical values and graphical representations, were not selected at random. The distractors reflect a faulty approach to reasoning that students might use and that might lead them astray. Instead of interpreting the mean CDPA score as a final measurement, we propose a different approach wherein we focus on the distribution of correct and incorrect responses. We use this distribution as an additional layer of analysis that can be derived when using this instrument. 
The original CDPA paper divides the survey by specific abilities and classifies them in two distinct groupings. The first grouping, composed of Q1, Q2, Q5, Q6, Q9, and Q10 \emph{evaluates students' understanding of the meaning and use of uncertainty in measurements}. The second grouping, Q3, Q4, Q7 and Q8, measures \emph{their ability to relate functions, graphs, and numbers}. In the analysis that follows, we chose to focus our attention on the first grouping since these questions specifically target the practical understanding of uncertainties. For further reference, a list of the ten questions, complete with their distractor answers and the justification for these answers, is available in the original paper \cite{day2011development}.

%\emph{Maybe not include this part?

\begin{figure*}[h!]
  \centering \includegraphics[width=0.9\textwidth]{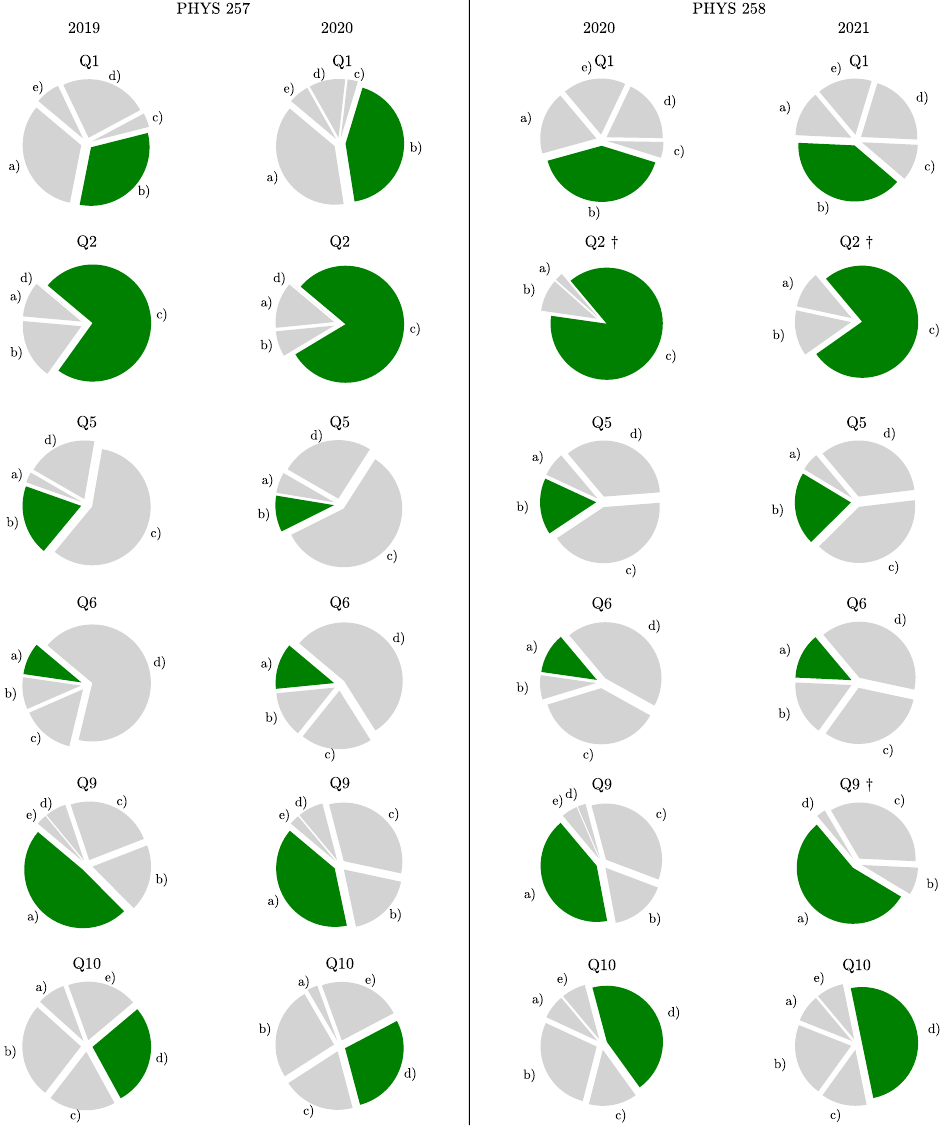}
  \caption{CDPA results for questions 1, 2, 5, 6, 9, and 10 for different years, and evolution as a function of level, comparing the post-test of PHYS 257 and PHYS 258. The correct multiple-choice answer is presented in green. The signs in Q2 and Q9 of PHYS 258 indicate no students chose alternatives d) in Q2 2020, and alternative e) in Q9 2021.}
  \label{fig:distribution_answers}
\end{figure*}

In Figure \ref{fig:distribution_answers}, we present the results for two complete years of undergraduate experimental physics laboratory education, PHYS 257 and PHYS 258. The pie charts show the correct answers highlighted in green and the incorrect answers are represented in gray. Remarkably, we notice that the two cohorts displayed in Figure \ref{fig:distribution_answers} show a very similar distribution of responses despite having taken the test in two different years. This trend becomes particularly evident when narrowing in on the correct responses given by respondents from both PHYS 257 and PHYS 258. When comparing the wrong alternative answers one by one, we also identify similarities that might point to distinct approaches to student reasoning. Moreover, in comparing the progression from one course to another, we notice a positive increase towards the correct answer for most questions.

We met with the professors to review the CDPA test questions. The professors agreed that the distractor responses indicated similar misconceptions to those described in the original CDPA paper. The professors evaluated each incorrect answer to identify possible misconceptions underlying the error. The professors were then asked to classify whether students should be able to answer each question by the end of the course they instructed. Their responses were categorized as follows: 1) yes, students should be able to answer the question, 2) no, students should not be able to answer the question, or 3) maybe, a stronger student in this course might be able to answer the question but it is not a requirement to pass the course in question. The analysis for each question, indicated in Table \ref{table:misconceptions} below, reflects the results of these feedback discussions. Overall, the mapping of misconceptions established by these McGill professors is in line with the original CDPA paper.

\begin{table*}[htb]
\centering
\footnotesize % You can use \footnotesize or \scriptsize for even smaller font
\resizebox{\textwidth}{!}{%
\begin{tabular}{c p{4cm} c c p{4cm}}
\hline
\textbf{Question} &
  \centering\arraybackslash\textbf{Content} &
  \centering\arraybackslash\textbf{\begin{tabular}[c]{@{}c@{}}Should students\\ be able to answer\\ in PHYS 257?\end{tabular}} &
  \centering\arraybackslash\textbf{\begin{tabular}[c]{@{}c@{}}Should students \\ be able to answer\\ in PHYS 258?\end{tabular}} &
  \centering\arraybackslash\textbf{\begin{tabular}[c]{@{}c@{}}Possible\\ misconception\end{tabular}} \\ \hline
Q1 &
  \begin{tabular}[c]{@{}p{4cm}@{}}Combining uncertainties \\ of different measurements\end{tabular} &
  M &
  Y &
  \begin{tabular}[c]{@{}p{4cm}@{}}Using weighted mean to \\ evaluate different uncertainties\end{tabular} \\ \hline
Q2 &
  Expressing significant figures &
  Y &
  Y &
  Displaying significant figures in text \\ \hline
Q5 &
  Identifying the best-fit curve &
  N &
  M &
  \begin{tabular}[c]{@{}p{4cm}@{}}Prioritizing more points rather\\ than the ones with smaller uncertainty \\ on a fit\end{tabular} \\ \hline
Q6 &
  Identifying the best-fit curve &
  N &
  M &
  \begin{tabular}[c]{@{}p{4cm}@{}}Prioritizing more points rather than the ones \\ with smaller uncertainty\\ on a fit\end{tabular} \\ \hline
Q9 &
  \begin{tabular}[c]{@{}p{4cm}@{}}Outcome of measurements and associated\\ instrumental uncertainties\end{tabular} &
  M &
  Y &
  \begin{tabular}[c]{@{}p{4cm}@{}}Evaluating different outcomes of measurements\\ and their individual influence on a series\end{tabular} \\ \hline
Q10 &
  \begin{tabular}[c]{@{}p{4cm}@{}}Combining uncertainties of \\ different measurements\end{tabular} &
  M &
  Y &
  \begin{tabular}[c]{@{}p{4cm}@{}}Uncertainty propagation when considering multiple\\ experiments\end{tabular} \\ \hline
\end{tabular}%
}\caption{After discussion with the professors teaching in the labs/classes, this table shows which of the CDPA questions students from PHYS 257 and PHYS 258 should be able to answer (Y), might be able to answer (M) and should not be able to answer (N), with a description of the possible misconception linked to an incorrect response.}
\label{table:misconceptions}
\end{table*}

\begin{figure*}[h!]
  \centering \includegraphics[width=\textwidth]{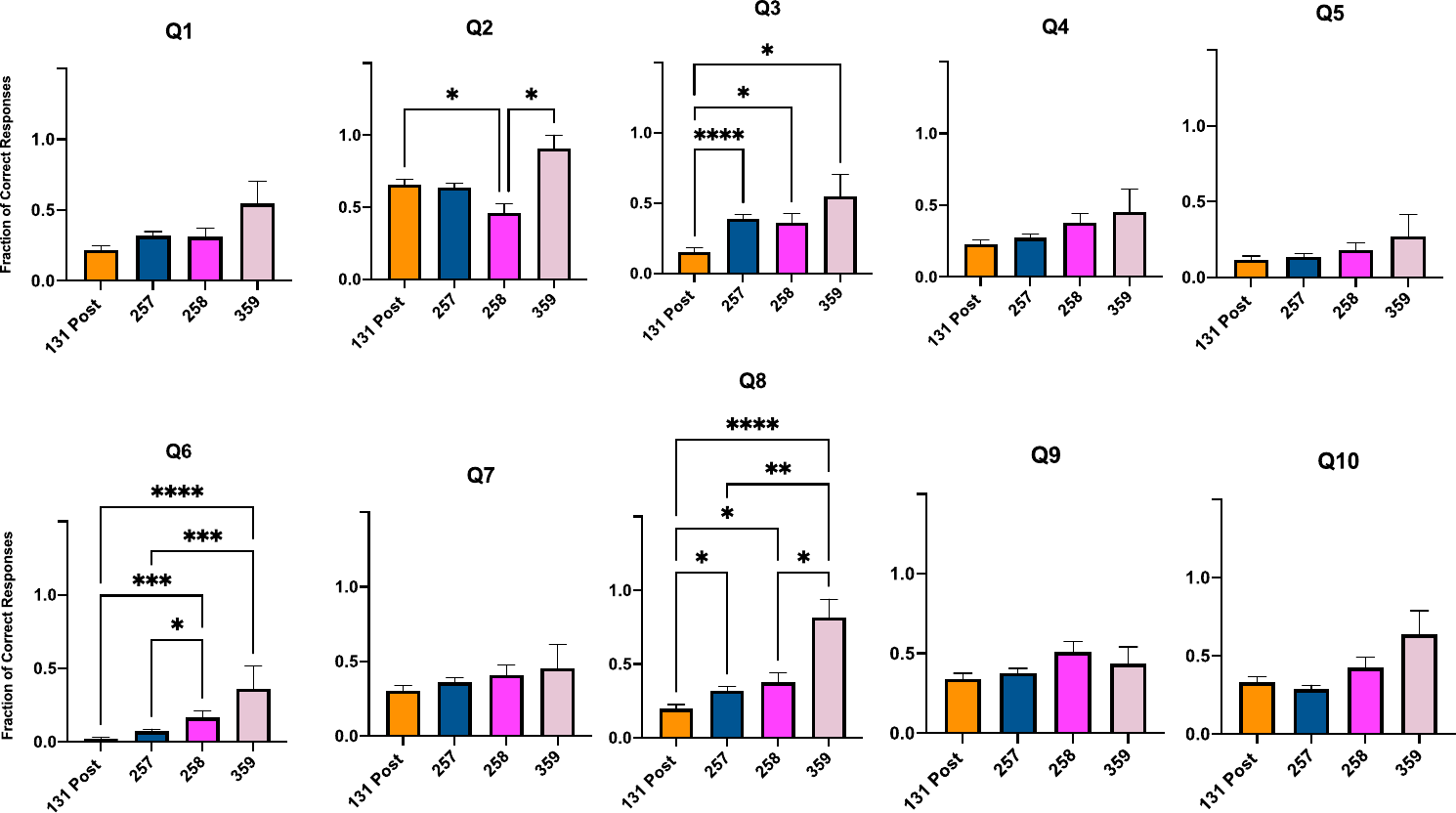}
  \caption{Distribution of correct responses for all the 10 CDPA questions for students in PHYS 131, PHYS 257, PHYS 258, and PHYS 359. The stars represent the degree of statistical significance in these results.}
  \label{fig:questions}
\end{figure*}

We present the results of individual questions in Figure \ref{fig:questions}. We notice an overall trend of increasing correct answers as students progress through their degree. To better understand these results, in Table \ref{table:misconceptions}, we show the possible misconceptions influencing the students' responses in the core experimental physics courses, focusing on PHYS 257 and PHYS 258. Students in their early years respond to Q1 incorrectly which is expected since they do not yet have the necessary skillset. The professors noted that the incorrect alternatives in Q1 show that students disregard uncertainties when taking the average of experimental data points. However they agreed that by the end of PHYS 258, students should be able to identify and treat different uncertainties in measurements correctly. 

Q2 received the highest rate of correct responses by students in all courses. This question addresses ability to understand significant figures in astronomical units. According to the professors, students should be able to respond to this question even before beginning the course. Incorrect alternatives do not present any specific insight into student reasoning other than a possible misconception about the expression of significant figures. 

In contrast, the professors pointed out that only students in PHYS 257 and PHYS 258 who are very strong should be able to answer Q5 and Q6. The concepts in these questions are introduced in PHYS 258, however students are not expected to compare different models for best fit until upper level courses. It must be reiterated that all the CDPA distractors are very close to the correct response and that, for these two questions, choosing an incorrect answer can be associated with disregarding numerical uncertainties graphically when considering a best fit line. For example, students often choose to fit the most points rather than fit the ones with the smallest uncertainty, an approach to reasoning that is indicative of their lack of knowledge on how to prioritize.

Q9 evaluates how students deal with the outcome of measurements and associated instrumental uncertainties. Students are not expected to answer this question correctly until they reach the end of PHYS 258. The distractors may reflect a misunderstanding of statistics and how statistics impact experimental measurements. 

Q10 investigates students' capacity to combine numerical uncertainties from different measurements. The professors expect students to master this critical ability by the end of PHYS 258, as the students develop a better understanding of how to handle uncertainties through the progression of experiments they conduct during the course, culminating in a final project they complete. This is reflected in the results when comparing PHYS 258 with earlier years. 

\subsection{What the CDPA reveals about upper level lab courses}
Having established a clearer picture of what the CDPA reveals about students' understanding of concepts related to uncertainty during their first comprehensive year of introductory lab courses, we now turn to what the CDPA can tell us about the level of learning that takes place in more advanced laboratory courses. As previously explained, beyond the 200-level courses, this study also includes data collected at the 300- and 400-levels, most notably PHYS 359 - Advanced Physics Laboratory 1, reserved for Honors Physics undergraduate students, and PHYS 439 - Majors Laboratory in Modern Physics, the equivalent course for Physics majors.

Between 2019 and 2022, the variability in CDPA score for PHYS 258 demonstrates that, at this point in their learning, students have not mastered the skills needed to evaluate experimental uncertainties. For PHYS 359, however, we note a significant increase in the CDPA score in 2022 that warrants a more in-depth investigation into what factors may have contributed to this leap in student understanding. As can be seen in Figure  \ref{fig:yearyear}, the 2022 mean CDPA score was significantly higher, from a class average of $2.5 \pm 2.0$ by the end of PHYS 258 to $4.5 \pm 1.0$ by the end of PHYS 359. Although we can expect noticeable improvement in the CDPA score after this upper-level laboratory course, the extent of this statistically significant increase over one semester left us with more questions.

Through discussion with the instructor who taught PHYS 359 during the Winter 2022 semester, we could ascertain that certain structural elements of the course contributed to this demonstrable increase in student understanding of uncertainties. Three key factors stand out in the course design and context that semester. First, the most significant difference between PHYS 258 and PHYS 359 is the class size. While PHYS 258 had over 100 students, the smaller enrollment numbers in PHYS 359 allowed for greater one on one interaction between students and their instructor. Second, students also spent more time in the lab each week, and third, they performed experiments that require a higher degree of precision and attention to uncertainties.

The PHYS 359 course has undergone changes and adaptations throughout the years. One example is that a course instructor for PHYS 359 in 2022 identified a noticeable weakness in incoming students’ level of understanding of uncertainties. This led to the development and implementation of an additional module dedicated to curve fitting in a self-paced Python tutorial. This module offered students more advanced exercises emphasizing graph interpretation and analysis, incorporating an interactive simulation feature and enabling personalized feedback. The teaching tool allowed the instructor to leverage the smaller class size for a more individualized instructional approach, dedicating time to helping students apply concepts related to uncertainties. With this personalized tool, the instructor could identify and correct student misconceptions more precisely.

Lastly, in addition to this novel instructional approach, the assessment scheme of PHYS 359 allocated 20 percent of the lab report grade, to skills such as \textit{justified quantitative statistical uncertainties, $\chi^2$ and residuals discussed quantitatively}. As a larger portion of the course grade depended on students demonstrating their mastery of these concepts, their understanding of uncertainties in applied contexts was further strengthened.

%We use the most important course and compare it in different scenarios, for the most important questions. We test them for consistency, and we show (pie charts) that they are consistent. 

%We test the 257 against 258 and identify some increase in the understanding, but still with underlying misconceptions (shown in the pie charts as well), that we can also discuss. This also makes sense because in the interviews with professors we also learn that students shouldn't be actually learning all of that in that level; they get exposure to the content, but it's not reinforced in the way they are graded (they haven't really had the need to use them, and don't really do that until 359). 

%We test against people below that level and above that level, and we show that there is a clear progression as well. Specifically in 359, there's an extra component that might help students overcome the misconceptions.

%Beyond that, we also have pandemic data to compare! We have results pre-pandemic, hybrid, and post-pandemic to compare. For Rebecca's reference: pedagogical impact of in-person experimentation.
%- ability mapping done with each professor, in each of those courses 
%- in terms of questions we get 1 correct, 6 wrong, 10 correct
%compare common misconceptions associated with each question to the professor notes as well
%We see some progression of student understanding between incoming 257-outgoing 258.
%Then, we show the 131 through 359 comparison

\section{Conclusions}

The findings presented in this study, which stemmed from a five year partnership between McGill University's Office of Science Education (OSE) and professors from the Physics Department, represent the longest running implementation of the CDPA test. As such, this research provides particular insight into how longitudinal research using the CDPA instrument can serve to support pedagogical research in individual laboratory courses and across departmental curricula. To our knowledge, this is the longest-running  CDPA survey, and the largest dataset, with a total of N=2023 students.

By comparing average CDPA scores from pre and post tests across all core laboratory courses in the McGill experimental physics undergraduate curriculum, we were able to distinguish distinct patterns in students’ understanding of concepts related to uncertainties from introductory 100-level courses through upper 300-level courses. As a general observation, our results show that, at the 100-level, students’ CDPA scores remained quite low and were mostly on-par with random guesswork. These results confirm UBC’s findings for introductory-level courses, which can be explained by the fact that, in first-year courses, students are introduced to the concept of uncertainties for the first time and are only expected to solve simple problems and conduct experiments without the need of deep conceptual understanding of uncertainties. 

Across all four years of this study, we noticed an increase in students' CDPA score from the 100 to the 200 level. This uptick in student understanding may be explained by the fact that at the 200-level, students complete a full academic year of laboratory courses, where they learn curve fitting, consideration of uncertainties, and production and interpretation of graphs. In PHYS 257 students are asked to assess systematic errors and perform error propagation and in PHYS 258 they need to consider statistical uncertainties to get the best-fit curves, understand and justify their reasoning, and differentiate between different models. In addition, PHYS 258 culminates in a small research project in which students are required to answer a research question of their choice, experimentally validate their measurements and demonstrate understanding of how the uncertainties should be considered in the analysis. Nevertheless, it is not until the 300-level and 400-level courses, when 20 percent of the lab report grade is allocated to justifying quantitative statistical uncertainties, that students’ demonstrated understanding of uncertainties develops into holistic concepts.
In addition to using the CDPA instrument as a mean to rank student understanding across the departmental laboratory curricula, our research also shows that this tool can be used to help professors comprehend the nature of their students’ misconceptions. As shown in the breakdown of CDPA incorrect responses, Figure \ref{fig:distribution_answers}, CDPA results can be used as a remediation tool to help professors identify gaps in student understanding, thereby providing an additional means to refine their teaching approach for subsequent cohorts. 

In conclusion, through this collaborative study, we were able to highlight new opportunities for pedagogical and structural improvement in physics laboratory courses. We show how a better understanding of students misconceptions can enhance professors' understanding of student learning on a course-by-course basis. This research reinforces the value of collecting program level data using a common instrument and then reporting the findings back to course professors for interpretation. It is important to note that engaging the professors in the process of data analysis, allowed us to pinpoint the educational benefit featured of this research. This study also points to a potential need for more inquiry-guided learning in laboratory courses, or, at the very least, it serves as an impetus to carve out more personalized attention in large-scale introductory laboratory courses. 

\section*{appendix}\label{appendixtables}

\subsection*{Table of collected data throughout the years}

All the collected data in this study is organized in Table \ref{tab:combined}, presented below. 
\begin{table*}[h!]
\centering
\begin{ruledtabular}
\begin{tabular}{lccccc}
\textbf{Class} & \textbf{Term} & \textbf{Format} & \textbf{Total} & \textbf{Tested} \\
\hline
\multirow{3}{*}{PHYS 101- Introduction to Mechanics} & Fall 2020 PRE  & Online  & 475 & 452 \\
                          & Fall 2021 PRE  & Online  & 77 & 34  \\
                          & Fall 2022 & In person  & 82 & 22 \\
\hline
\textbf{Total Tested (PHYS 101)} & & & & \textbf{N = 508} \\
\hline
\multirow{2}{*}{PHYS 131- Mechanics and Waves} & Fall 2019 PRE  & In person  & 177 & 167 \\
                          & Fall 2019 POST & In person  & 323 & 313 \\
                          & Fall 2020 PRE & Online & 335 & 320 \\
                          & Fall 2020 POST & Online & 191 & 186 \\
                          & Fall 2021 & Online & 78 & 37 \\
                          & Fall 2020 & Online & 52 & 32 \\
\hline
\textbf{Total Tested (PHYS 131)} & & & & \textbf{N = 1055} \\
\hline
\multirow{4}{*}{PHYS 257- Experimental Methods I} & Fall 2019   & In person            & 105 & 104 \\
                          & Fall 2020   & Online               & 74  & 71  \\
                          & Fall 2021   & In person and online & 65  & 64  \\
                          & Fall 2022   & In person            & 50  & 43  \\
\hline
\textbf{Total Tested (PHYS 257)} & & & & \textbf{N = 282} \\
\hline
\multirow{4}{*}{PHYS 258- Experimental Methods II} & Winter 2020 & In person and online & 44  & 44  \\
                          & Winter 2021 & Online               & 39  & 38  \\
                          & Winter 2022 & In person and online & 44  & 40  \\
                          & Winter 2023 & In person            & 23  & 19  \\
\hline
    \textbf{Total Tested (PHYS 258)} & & & & \textbf{N = 141} \\
\hline
\multirow{1}{*}{PHYS 359 - Honors Laboratory in Modern Physics I} & 2023       & In person & 23 & 19 \\
\hline
\textbf{Total Tested (PHYS 359)} & & & & \textbf{N = 19} \\
\hline
\multirow{1}{*}{PHYS 434- Optics} & Fall 2020  & Online    & 17 & 16 \\
\hline
\textbf{Total Tested (PHYS 434)} & & & & \textbf{N = 16} \\
\hline
\multirow{1}{*}{PHYS 469-Honors Laboratory  in Modern  Physics 2} & Fall 2020  & Online    & 2  & 2  \\
\hline
\textbf{Total Tested (PHYS 469)} & & & & \textbf{N = 2} \\
\end{tabular}
\end{ruledtabular}
\caption{Summary of class formats, and testing for the various physics courses tested using the CDPA across different terms. The total refers to the number of students who answered the survey in that specific year, willing to participate or not in the survey.}
\label{tab:combined}
\end{table*}

\clearpage

\bibliography{apssamp}% Produces the bibliography via BibTeX.

\end{document}